\documentclass[aps,prb,amsmath,twocolumn]{revtex4}
\usepackage{bm}
\usepackage{graphicx}

\DeclareMathOperator{\sgn}{sgn}
\DeclareMathOperator{\Img}{\mathrm{Im}}
\DeclareMathOperator{\Rea}{\mathrm{Re}}
\DeclareMathOperator{\Sp}{\mathrm{Sp}}

\begin{document}
\title{Josephson current in ballistic heterostructures with spin-active interfaces}
\author{Mikhail S. Kalenkov}
\affiliation{I.E. Tamm Department of Theoretical Physics, P.N.
Lebedev Physics Institute, 119991 Moscow, Russia}
\author{Artem V. Galaktionov}
\affiliation{I.E. Tamm Department of Theoretical Physics, P.N.
Lebedev Physics Institute, 119991 Moscow, Russia}
\author{Andrei D. Zaikin}
\affiliation{Forschungszentrum Karlsruhe, Institut f\"ur Nanotechnologie,
76021, Karlsruhe, Germany}
\affiliation{I.E. Tamm Department of Theoretical Physics, P.N.
Lebedev Physics Institute, 119991 Moscow, Russia}

\begin{abstract}
We develop a general microscopic theory of dc Josephson effect
in hybrid SNS structures with ballistic electrodes and spin-active NS
interfaces. We establish a direct relation between the spectrum of Andreev
levels and the Josephson current which contains complete information
about non-trivial interplay between Andreev reflection and spin-dependent interface
scattering. The system exhibits a rich structure of
properties  sensitive to spin-dependent barrier transmissions,
spin-mixing angles, relative magnetization orientation of interfaces and the
kinematic phase of scattered electrons. We analyze the current-phase relations
and identify the conditions for the presence of a $\pi$-junction state in the
systems under consideration. We also analyze resonant enhancement of the
supercurrent in gate-voltage-driven nanojunctions. As compared to the
non-magnetic case, this effect can be strongly modified by spin-dependent
scattering at NS interfaces.

\end{abstract}

\maketitle

\section{Introduction}
Spin-sensitive Andreev reflection in superconducting hybrid
structures yields a number of interesting and non-trivial effects
which have been addressed and studied in the literature. Already more
than three decades ago it was realized that spin-flip electron
tunneling through a magnetic interface between two superconductors
may cause a sign change of the Josephson current and yield the
so-called $\pi$-junction state in superconducting weak links
\cite{Leva}. The same effect can also occur in
superconductor-ferromagnet-superconductor (SFS) junctions where
the Josephson critical current was predicted to oscillate as a
function of the ferromagnet layer thickness also leading to
formation of the $\pi$-junction state \cite{BBP,GKI}. These
theoretical predictions were confirmed  in experiments with SFS
junctions \cite{Valera,Kontos}.

More recently it was realized \cite{BVE,KSJ} that the Cooper pair
wave function in SF hybrids may change its symmetry from a singlet
in a superconductor to a triplet in a ferromagnet. This so-called
odd-frequency pairing state \cite{BVE} implies that Cooper pairs
can penetrate deep into the ferromagnet thus causing the long
range proximity effect in SF systems. Experimental evidence for
such long range coherent behavior of SF hybrids was discussed in
Ref. \onlinecite{Petrashov}.

Another interesting realization of this long range proximity
effect is the possibility for non-vanishing Josephson current to
flow across superconducting weak links containing strong
ferromagnets or the so-called half-metals (H). Note that
spin-singlet Cooper pairs cannot penetrate into H-metals because
such metals are fully spin polarized materials acting as
insulators for electrons with one of the two spin directions.
Hence, no supercurrent carried by spin-singlet Cooper pairs can
occur in SHS junctions. Superconducting correlations can
nevertheless survive deep inside strong ferromagnets provided
there exists a mechanism for spin-flip scattering at both HS
interfaces of the junction. This mechanism allows for conversion
of spin-singlet pairing in S-electrodes into spin-triplet pairing
in H-metals thus making it possible for the supercurrent to flow
across the system. A theory of this non-trivial Josephson effect
in SHS junctions was recently addressed by a number of authors
\cite{Eschrig,Sasha,GKZ08}. Experimental results \cite{Keizer}
appear to support that non-vanishing supercurrent can indeed flow
across sufficiently thick SHS junctions.

The main goal of this paper is to develop a general theory of dc
Josephson effect in SNS heterostructures with spin-active
interfaces. Previously \cite{KZ07} we already demonstrated that
spin-dependent scattering at such interfaces yields a number of
interesting and non-trivial properties of non-local Andreev
reflection in three-terminal NSN devices. Here we find that both
the Josephson critical current and the current-phase relation in
SNS junctions are very sensitive to particular values of ($a$)
spin-dependent interface transmissions, ($b$) spin-mixing angles
and ($c$) the electron kinematic phase showing a rich variety of
features which can be detected and studied in future experiments.

One possible experimental realization of SNS junctions with
spin-active interfaces is achieved by placing a thin layer of some
magnetic material at the interfaces between superconducting and
normal metals. In this case transmission probabilities for
spin-up and spin-down electrons propagating through such
interfaces may take different values. In addition, the scattering
phase of incoming electrons may also depend on their spin states.
This physical situation can be modelled by a spin-active interface
described by two (spin-up and spin-down) transmission
probabilities and by the so-called spin-mixing angle which is just
the difference between the scattering phases for the spin-up and
spin-down states of incoming electrons. Yet one more parameter --
the kinematic phase -- should be introduced in order to account
for the phase acquired by an electron between successive
scattering events at NS interfaces. This phase is essentially set
by the product of the Fermi momentum and the distance covered by
an electron between two scattering events. Since the electron
momentum can be controlled (shifted) by applying the gate voltage
the system can be periodically driven to and out of resonance
thereby rendering a possibility to experimentally investigate the
dependence of the Josephson current on the kinematic phase.

The structure of the paper is as follows. In Sec. 2 we will define
our model and specify some key equations that will be employed in
our further consideration. Sec. 3 is devoted to the analysis of
Andreev bound states in SNS junctions with spin-active interfaces.
In Sec. 4 we will demonstrate that the Josephson current in our
structure can be directly expressed in terms of the Andreev
spectrum and derive the general expression for this current. This
expression will then be analyzed in Sec. 5 in a number of
interesting limits. In Sec. 6 we will briefly summarize our main
observations. Some technical details of our analysis will be
presented in Appendices A and B.

\section{The model and basic equations}
\begin{figure}
\centerline{\includegraphics{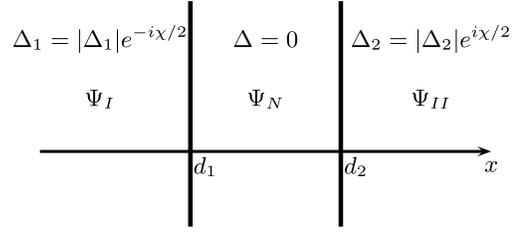}} \caption{SNS junction
with spin-active interfaces located at $x=d_1$ and $x=d_2$. }
\label{sfnfs-fig}
\end{figure}

For our analysis we will employ the standard model of a planar
ballistic SNS junction (Fig. \ref{sfnfs-fig}) with spin-active NS
interfaces (located at $x=d_1$ and $x=d_2$) and cross-section area
$\mathcal{A}$. Outside an immediate vicinity of the interface
regions quasiparticle wave functions can be represented as a sum
of two rapidly oscillating exponents
\begin{equation}
\Psi= \Psi_+(x) e^{ip_{Fx}x +i\bm{p}_{\parallel} \bm{\rho}} +
\Psi_-(x) e^{-ip_{Fx}x+ i\bm{p}_{\parallel} \bm{\rho}},
\end{equation}
where $x$ is coordinate normal to the NS interfaces, $\bm{\rho}$
represents the coordinates in the transversal directions and
$p_{Fx}=\sqrt{p_F^2-\bm{p}_{\parallel}^2}>0$ is the normal
component of the Fermi momentum. The wave functions
$\Psi_{\pm}(x)$ vary smoothly at atomic distances. Applying the
standard Andreev approximation it is easy to demonstrate that
$\Psi_{\pm}(x)$  obeys the following equation
\begin{equation}
\left[
\varepsilon \pm i\tau_3 v_{Fx}\dfrac{\partial}{\partial x}
-
\begin{pmatrix}
0 & i\sigma_2 \Delta(x) \\
-i\sigma_2 \Delta(x)^* & 0 \\
\end{pmatrix}
\right]
\Psi_{\pm}=0,
\label{andreeveq}
\end{equation}
where $\hat\tau_3$ is the Pauli matrix in the Nambu space,
$\sigma_i$ are the Pauli matrices in the spin space, $v_{Fx}=p_{Fx}/m$
and $\Delta$ is the BCS order parameter which is assumed to be
spatially constant equal to $|\Delta_1|e^{-i\chi/2}$
($|\Delta_2|e^{i\chi/2}$) in the first (second) superconducting
electrode and is set equal to zero $\Delta =0$ in the normal
metal. Eq. \eqref{andreeveq} does not apply at the interfaces and
should be supplemented by the proper boundary conditions which
account for spin-sensitive electron scattering. In order to
formulate these boundary conditions we will employ the general
$S$-matrix formalism. For instance, matching of the quasiparticle
wave functions at the first interface is performed with the aid of
the following equation
\begin{equation}
\begin{pmatrix}
\Psi_{1-}\\
\Psi_{1'+}
\end{pmatrix}
=
\begin{pmatrix}
\hat S_{11} & \hat S_{11'} \\
\hat S_{1'1} & \hat S_{1'1'} \\
\end{pmatrix}
\begin{pmatrix}
\Psi_{1+}\\
\Psi_{1'-}
\end{pmatrix}.
\label{bcond}
\end{equation}
Here  $\Psi_{1+}$, $\Psi_{1-}$ and $\Psi_{1'+}$, $\Psi_{1'-}$ are
the quasiparticle wave function amplitudes respectively at
superconducting and normal sides of the first interface. The
matrices $\hat S_{ij}$ are diagonal in the Nambu space
\begin{equation}
\hat S_{ij}
=
\begin{pmatrix}
S_{ij} & 0 \\
0 & \underline{S}_{ij} \\
\end{pmatrix},
\end{equation}
where $S_{ij}$ and $\underline{S}_{ij}$ are the building blocks of
the full electron and hole interface S-matrices\cite{Millis88}
\begin{gather}
\mathcal{S}_1=
\begin{pmatrix}
S_{11} & S_{11'} \\
S_{1'1} & S_{1'1'} \\
\end{pmatrix}, \quad
\underline{\mathcal{S}}_1=
\begin{pmatrix}
\underline{S}_{11} & \underline{S}_{11'} \\
\underline{S}_{1'1} & \underline{S}_{1'1'} \\
\end{pmatrix},
\label{matr}\\
\mathcal{S}_1\mathcal{S}_1^+=1, \quad \underline{\mathcal{S}}_1
\underline{\mathcal{S}}_1^+=1. \label{matr2}
\end{gather}
In general electron and hole S-matrices do not coincide with each
other but obey the following relation
\begin{equation}
\underline{\mathcal{S}}_1(\bm{p}_{\parallel})=\mathcal{S}_1^T(-\bm{p}_{\parallel}).
\label{srel}
\end{equation}

Electron scattering at the second interface is described
analogously with the aid of the matrices $\mathcal{S}_2$
and $\underline{\mathcal{S}}_2$ which have the same structure as
in Eqs. (\ref{matr})-(\ref{srel}).

\section{Andreev states}

With aid of the above equations it is convenient to analyze the
spectrum of Andreev bound states in SNS heterostuctures with spin
active interfaces. For this purpose it is necessary to solve Eq.
\eqref{andreeveq} and explicitly find quasiparticle wave
functions of our system. In the first superconductor $x<d_1$ the
wave function has the form
\begin{multline}
\Psi_{I}=
\begin{pmatrix}
-i\sigma_2 e^{-i\chi/2} a_1 A_1\\
A_1
\end{pmatrix}
e^{ip_{Fx}(x-d_1)}
e^{\kappa_1(x-d_1)}
+\\+
\begin{pmatrix}
 A_2\\
i\sigma_2 e^{i\chi/2} a_1 A_2
\end{pmatrix}
e^{-ip_{Fx}(x-d_1)}
e^{\kappa_1(x-d_1)},
\end{multline}
where $\kappa_m=\sqrt{|\Delta_m|^2-\varepsilon^2}/v_{Fx}$ and $a_m
= ( -\varepsilon + \sqrt{ \varepsilon^2 - |\Delta_m|^2}
)/|\Delta_m|$ ($m=1,2$). The functions $a_m(\varepsilon)$ are
analytic in the upper half-plane of the complex variable
$\varepsilon$. For real $\varepsilon$ they read
\begin{equation}
a_m(\varepsilon)=
\begin{cases}
\dfrac{-\varepsilon + i
\sqrt{|\Delta_m|^2-\varepsilon^2}}{|\Delta_m|}, &
|\varepsilon|<|\Delta_m|,
\\
\dfrac{-\varepsilon +  \sqrt{\varepsilon^2 -|\Delta_m|^2}
\sgn(\varepsilon)}{|\Delta_m|}, & |\varepsilon|>|\Delta_m|,
\end{cases}
\end{equation}
while for purely imaginary values of $\varepsilon$ we have
\begin{equation}
a_m(i\omega)=i\dfrac{\sqrt{\omega^2+|\Delta_m|^2}-\omega}{|\Delta_m|},
\quad \omega > 0.
\end{equation}

In the normal metal $d_1<x<d_2$ and in the second superconductor
$x>d_2$ one has respectively
\begin{multline}
\Psi_{N}=
\begin{pmatrix}
e^{i\varepsilon(x-d_1)/v_{Fx}} B_1 \\
e^{-i\varepsilon(x-d_1)/v_{Fx}} B_2 \\
\end{pmatrix}
e^{ip_{Fx}(x-d_1)}
+\\+
\begin{pmatrix}
e^{-i\varepsilon(x-d_1)/v_{Fx}} B_3 \\
e^{i\varepsilon(x-d_1)/v_{Fx}} B_4 \\
\end{pmatrix}
e^{-ip_{Fx}(x-d_1)}
\end{multline}
and
\begin{multline}
\Psi_{II}=
\begin{pmatrix}
C_1\\
i\sigma_2 e^{-i\chi/2} a_2 C_1
\end{pmatrix}
e^{ip_{Fx}(x-d_2)}
e^{-\kappa_2(x-d_2)}
+\\+
\begin{pmatrix}
-i\sigma_2 e^{ i\chi/2} a_2 C_2\\
C_2
\end{pmatrix}
e^{-ip_{Fx}(x-d_2)}
e^{-\kappa_2(x-d_2)}.
\end{multline}
All coefficients $A_{1,2}$, $B_{1,2,3,4}$ and $C_{1,2}$ are
vectors in the spin space which should be determined from the
matching conditions at both NS interfaces. From Eq. (\ref{bcond})
at the first interface ($x=d_1$) we obtain
\begin{gather}
A_2 =
S_{11} (-i\sigma_2 a_1 e^{- i\chi/2})A_1 + S_{11'} B_3,
\label{eq1}
\\
(i\sigma_2 a_1 e^{ i\chi/2})A_1=\underline{S}_{11} A_1 + \underline{S}_{11'}^+ B_4,
\\
B_1=S_{1'1} (-i\sigma_2 a_1 e^{- i\chi/2}) A_1 + S_{1'1'} B_3,
\\
B_2=\underline{S}_{1'1} A_1 + \underline{S}_{1'1'} B_4.
\end{gather}
These equations yield the relation between the amplitudes for the
incoming and outgoing electron and hole waves (respectively $B_3$,
$B_4$ and $B_1$, $B_2$):
\begin{equation}
\begin{pmatrix}
B_1 \\B_2 \\
\end{pmatrix}
= K_1
\begin{pmatrix}
B_3 \\B_4 \\
\end{pmatrix},
\label{k1rel}
\end{equation}
where $K_1$ is the scattering matrix which describes both normal
and Andreev reflection at the first interface. This matrix has the
form
\begin{widetext}
\begin{equation}
K_1=
\begin{pmatrix}
S_{1'1'} + a_1^2 S_{1'1}\sigma_2 (\underline{S}_{11}-a_1^2 \sigma_2 S_{11} \sigma_2 )^{-1} \sigma_2 S_{11'} &
i a_1 e^{-i\chi/2} S_{1'1} \sigma_2 (\underline{S}_{11}-a_1^2 \sigma_2 S_{11} \sigma_2 )^{-1} \underline{S}_{11'} \\
i a_1 e^{i\chi/2} \underline{S}_{1'1} (\underline{S}_{11}-a_1^2 \sigma_2 S_{11} \sigma_2 )^{-1} \sigma_2 S_{11'} &
\underline{S}_{1'1'}- \underline{S}_{1'1} (\underline{S}_{11}-a_1^2 \sigma_2 S_{11} \sigma_2 )^{-1} \underline{S}_{11'}\\
\end{pmatrix}
\label{k1}
\end{equation}
\end{widetext}
Diagonal blocks of this matrix describes electron-electron and
hole-hole amplitudes while off-diagonal blocks account for
electron-hole conversion. One can directly verify that the matrix
$K_1$ is unitary for subgap energies $|\varepsilon| < |\Delta_1|$
at which no excitations exist in the superconducting electrode I.

Analogously from the matching conditions at the second interface
($x=d_2$)  we obtain
\begin{equation}
\begin{pmatrix}
B_3 \\B_4 \\
\end{pmatrix}
= e^{-i\varphi}Q K_2 Q
\begin{pmatrix}
B_1 \\B_2 \\
\end{pmatrix},
\label{k2rel}
\end{equation}
where $K_2$ is scattering matrix for the second interface at
subgap energies $|\varepsilon| < |\Delta_2|$ and we introduced the
following notations
\begin{equation}
e^{2ip_{Fx}(d_2-d_1)}=e^{-i\varphi}, \quad
e^{-i\varepsilon(d_2-d_1)/v_{Fx}}=q^2.
\label{qdef}
\end{equation}
The matrix $K_2$ is defined by Eq. \eqref{k1} with $1\rightarrow
2$ and $\chi\leftrightarrow -\chi$. The transfer matrix $Q$ has a
simple diagonal structure
\begin{equation}
Q=
\begin{pmatrix}
q^{-2}\sigma_0 & 0 \\
0 & q^2\sigma_0 \\
\end{pmatrix}.
\end{equation}
It provides the relations between $\Psi_{\pm}$-amplitudes at both
NS interfaces (on the normal metal side):
\begin{gather}
\Psi_{N+}(d_2)=Q^{-1}\Psi_{N+}(d_1),
\\
\Psi_{N-}(d_2)=Q\Psi_{N-}(d_1).
\end{gather}

Combining Eqs. \eqref{k1rel} and \eqref{k2rel} we obtain the
following general condition which defines the energies of Andreev
bound states in our system:
\begin{equation}
P(\varepsilon, \bm{p}_{\parallel},\chi)=
\det\left| 1-e^{-i\varphi}Q  K_2 Q  K_1  \right|=0
\label{peq}
\end{equation}
Note that scattering properties of the first (second) interface
enter into $P$ only through the matrix $K_1$ ($K_2$) while the
transfer matrix $Q$ and the kinematic phase $\varphi$ determine
the dependence of the bound state energy on the thickness of the
normal metal layer $d=d_2-d_1$. It is also worth pointing out that
our Eq. \eqref{peq} can be rewritten in terms of the effective
energy dependent scattering matrix \cite{Beenakker91,Beenakker08}
\begin{multline}
P_1(\varepsilon, \bm{p}_{\parallel},\chi)=\\=
\det \left|
\underline{\mathcal{S}}(\bm{p}_{\parallel},\varepsilon) -
A(\varepsilon,\chi)
\mathcal{S}(\bm{p}_{\parallel},\varepsilon)
A(\varepsilon,-\chi)
\right|=0,
\end{multline}
where
\begin{equation}
A(\varepsilon,\chi)=
\begin{pmatrix}
a_1\sigma_2 e^{i\chi/2} & 0 \\
0 & a_2\sigma_2 e^{-i\chi/2}\
\end{pmatrix},
\end{equation}
and $\mathcal{S}(\bm{p}_{\parallel},\varepsilon)$,
$\underline{\mathcal{S}}(\bm{p}_{\parallel},\varepsilon)$ are
effective electron and hole scattering matrices
\begin{gather}
\mathcal{S}(\bm{p}_{\parallel},\varepsilon)=
\begin{pmatrix}
\mathcal{S}_{11}(\bm{p}_{\parallel},\varepsilon) &
\mathcal{S}_{12}(\bm{p}_{\parallel},\varepsilon) \\
\mathcal{S}_{21}(\bm{p}_{\parallel},\varepsilon) &
\mathcal{S}_{22}(\bm{p}_{\parallel},\varepsilon) \\
\end{pmatrix},
\\
\underline{\mathcal{S}}(\bm{p}_{\parallel},\varepsilon)
=\mathcal{S}^T(-\bm{p}_{\parallel},-\varepsilon),
\end{gather}
\begin{widetext}
\begin{gather}
\mathcal{S}_{11}(\bm{p}_{\parallel},\varepsilon)=
S_{11} + q^{-4} e^{-i\varphi}S_{11'}
\left(1-q^{-4} e^{-i\varphi} S_{2'2'} S_{1'1'} \right)^{-1} S_{2'2'} S_{1'1},
\\
\mathcal{S}_{12}(\bm{p}_{\parallel},\varepsilon)=
q^{-2} e^{-i\varphi/2} S_{11'}
\left(1-q^{-4} e^{-i\varphi} S_{2'2'} S_{1'1'}\right)^{-1} S_{2'2},
\\
\mathcal{S}_{21}(\bm{p}_{\parallel},\varepsilon)=
q^{-2} e^{-i\varphi/2} S_{22'}
\left(1-q^{-4} e^{-i\varphi} S_{1'1'} S_{2'2'}\right)^{-1} S_{1'1},
\\
\mathcal{S}_{22}(\bm{p}_{\parallel},\varepsilon)=
S_{22} + q^{-4} e^{-i\varphi}  S_{22'}
\left(1-q^{-4} e^{-i\varphi} S_{1'1'} S_{2'2'}\right)^{-1}   S_{1'1'} S_{2'2}
\end{gather}
\end{widetext}
The functions $P(\varepsilon, \bm{p}_{\parallel},\chi)$ and
$P_1(\varepsilon, \bm{p}_{\parallel},\chi)$ are proportional to
each other with the $\chi$-independent proportionality factor. The
matrices $\mathcal{S}(\bm{p}_{\parallel},\varepsilon)$ and
$\underline{\mathcal{S}}(\bm{p}_{\parallel},\varepsilon)$ describe
normal state properties of our device while superconductivity is
accounted for by the matrix $A(\varepsilon,\chi)$ which depends on
superconducting order parameters $\Delta_{1,2}$ and the phase
difference $\chi$.

\section{Josephson current: General expressions}
It turns out that the function $P(\varepsilon,
\bm{p}_{\parallel},\chi)$ (\ref{peq}) can be directly used in
order to derive a general and compact expression for the Josephson
current across SNS structure under consideration. In order to accomplish this
task we will use the standard Green function formalism \cite{AGD}.
The corresponding derivation is presented in Appendix A. It
yields the following result
\begin{equation}
I(\chi)=
-2 e \mathcal{A} T
\sum_{\omega_n>0}
\int\limits_{|p_\parallel|<p_F}
\frac{d^2 p_\parallel}{(2\pi)^2}
\dfrac{\partial P(i\omega_n, \bm{p}_{\parallel},\chi)/\partial\chi
}{P(i\omega_n, \bm{p}_{\parallel},\chi)},
\label{Ieq}
\end{equation}
where $\omega_n = \pi T (2n+1)$ is Matsubara frequency. Using
standard transformation summation over Matsubara frequencies in
Eq. \eqref{Ieq} can be rewritten in terms of the integral
\begin{multline}
I(\chi)=
-\dfrac{ e \mathcal{A}}{2\pi}
\int d\varepsilon \tanh\dfrac{\varepsilon}{2T}
\times\\\times
\int\limits_{|p_\parallel|<p_F}
\frac{d^2 p_\parallel}{(2\pi)^2}
\Img
\dfrac{\partial P(\varepsilon, \bm{p}_{\parallel},\chi)/\partial\chi
}{P(\varepsilon+i0, \bm{p}_{\parallel},\chi)}.
\label{Ieq1}
\end{multline}
At energies close to the bound state $P$ becomes a linear function
of the energy variable, $P(\varepsilon, \bm{p}_{\parallel}, \chi )
\propto (\varepsilon-\varepsilon_B(\bm{p}_{\parallel}, \chi))$.
With this in mind one can easy check that Eq. \eqref{Ieq1}
correctly describes the contribution from Andreev bound states. In
Appendix \ref{green} we also demonstrate that Eqs. \eqref{Ieq} and
\eqref{Ieq1} properly account for the scattering states
contribution to $I(\chi )$. We can also add that momentum
integration in Eqs. \eqref{Ieq}, \eqref{Ieq1} can be transformed
into the sum over conducting channels
\begin{equation}
\mathcal{A}
\int\limits_{|p_\parallel|<p_F}\frac{d^2p_\parallel}{(2\pi)^2}
\rightarrow \sum_k .
\label{chansum}
\end{equation}

The above expressions for $I(\chi )$ define the Josephson current
in SNS junctions with spin-active interfaces under very general
conditions. In various physical situations underlying symmetries
of the model typically limit the number of parameters of the
scattering matrices ${\cal S}_{1,2}$. One important example was
considered in Ref. \onlinecite{GKZ08} where we analyzed the
Josephson current across a half-metallic layer in-between two
superconducting electrodes. In that case one needs to assume that
spin-active interfaces do not possess inversion symmetry
\cite{Eschrig,GKZ08} thus allowing for spin-flip scattering.
Substituting the particular form of the scattering matrix
\cite{GKZ08} into the general expressions for $I(\chi )$ derived
here one immediately recovers the results \cite{GKZ08}.

In this paper we consider another generic type of the scattering
matrix describing spin-active interfaces which do possess
inversion symmetry as well as reflection symmetry in the plane
normal to the corresponding interface. Such $\mathcal{S}$-matrices
can be chosen in the following form
\begin{multline}
S_{11}=S_{1'1'}=\underline{S}^T_{11}=\underline{S}^T_{1'1'}
=\\
U(\alpha)
\begin{pmatrix}
\sqrt{R_{1\uparrow}}e^{i\theta_1/2} & 0 \\
0 & \sqrt{R_{1\downarrow}}e^{-i\theta_1/2} \\
\end{pmatrix}e^{i\zeta_1}
U^+(\alpha),
\label{S11}
\end{multline}
\begin{multline}
S_{11'}=S_{1'1}=\underline{S}^T_{11'}=\underline{S}^T_{1'1}
=\\
U(\alpha) i
\begin{pmatrix}
\sqrt{D_{1\uparrow}}e^{i\theta_1/2} & 0 \\
0 & \sqrt{D_{1\downarrow}}e^{-i\theta_1/2} \\
\end{pmatrix}e^{i\zeta_1}
U^+(\alpha),
\label{S11'}
\end{multline}
\begin{multline}
S_{22}=S_{2'2'}=\underline{S}_{22}=\underline{S}_{2'2'}
=\\
\begin{pmatrix}
\sqrt{R_{2\uparrow}}e^{i\theta_2/2} & 0 \\
0 & \sqrt{R_{2\downarrow}}e^{-i\theta_2/2} \\
\end{pmatrix}e^{i\zeta_2},
\label{S22}
\end{multline}
\begin{multline}
S_{22'}=S_{2'2}=\underline{S}_{22'}=\underline{S}_{2'2}
=\\
i
\begin{pmatrix}
\sqrt{D_{2\uparrow}}e^{i\theta_2/2} & 0 \\
0 & \sqrt{D_{2\downarrow}}e^{-i\theta_2/2} \\
\end{pmatrix}e^{i\zeta_2}.
\label{S22'}
\end{multline}
Here $R_{1(2)\uparrow (\downarrow )}=1-D_{1(2)\uparrow (\downarrow
)}$ are spin dependent reflection coefficients of both NS
interfaces, $\theta_{1,2}$ are spin-mixing angles, $\zeta_{1,2}$
are real phase parameters  and $U(\alpha)$ is the rotation matrix
in the spin space which depends on the angle $\alpha$ between
polarizations of the two interfaces,
\begin{equation}
U(\alpha)=\exp(-i\alpha\sigma_1/2)=
\begin{pmatrix}
\cos( \alpha/2) & -i \sin(\alpha/2) \\
-i \sin(\alpha/2) & \cos( \alpha/2) \\
\end{pmatrix}.
\label{pma}
\end{equation}
We would like to emphasize that -- in contrast to the case studied
in Refs. \onlinecite{Eschrig,GKZ08} -- the scattering matrices
(\ref{S11})-(\ref{pma}) conserve the projection of the electron
spin along the interface polarization axis and {\it do not} allow
for spin-flip scattering.

Let us evaluate both the function $P(\varepsilon,
\bm{p}_{\parallel},\chi)$ and the Josephson current in the case of
non-collinear barriers polarizations. Absorbing the phases
$\zeta_{1,2}$ into the kinematic phase
\begin{equation}
\varphi=-2p_{Fx}d-\zeta_1-\zeta_2,
\end{equation}
after some algebra we obtain
\begin{widetext}
\begin{multline}
P(\varepsilon, \bm{p}_{\parallel},\chi)= \Biggl\{ \left[\cos\chi +
W(D_{1\uparrow} , D_{1\downarrow} , \theta_1 , D_{2\uparrow} ,
D_{2\downarrow} , \theta_2, \varepsilon)\right] \left[\cos\chi +
W^*(D_{1\uparrow} , D_{1\downarrow} , \theta_1 , D_{2\uparrow} ,
D_{2\downarrow} , \theta_2, -\varepsilon)\right]\cos^2(\alpha/2)
+\\+ \left[\cos\chi + W(D_{1\downarrow}, D_{1\uparrow}, -\theta_1
, D_{2\uparrow} , D_{2\downarrow} , \theta_2, \varepsilon)\right]
\left[\cos\chi + W^*( D_{1\downarrow}, D_{1\uparrow} , -\theta_1 ,
D_{2\uparrow} , D_{2\downarrow} , \theta_2,
-\varepsilon)\right]\sin^2(\alpha/2) -\\-
\sin^2(\alpha/2)\cos^2(\alpha/2) \left(
\left|\sqrt{R_{1\uparrow}}e^{i\theta_1/2}-\sqrt{R_{1\downarrow}}e^{-i\theta_1/2}\right|^2
(1+a_1^2)^2 -4a_1^2\sin^2\theta_1\right) \times\\\times \left(
\left|\sqrt{R_{2\uparrow}}e^{i\theta_2/2}-\sqrt{R_{2\downarrow}}e^{-i\theta_2/2}\right|^2
(1+a_2^2)^2 -4a_2^2\sin^2\theta_2\right) \dfrac{1}{4a^4
D_{1\uparrow}D_{1\downarrow}D_{2\uparrow}D_{2\downarrow}}
\Biggr\}, \label{PPP}
\end{multline}
where
\begin{multline}
W(D_{1\uparrow} , D_{1\downarrow} , \theta_1 , D_{2\uparrow} ,
D_{2\downarrow} , \theta_2, \varepsilon)= -\dfrac{1}{2a_1 a_2
\sqrt{D_{1\uparrow}D_{1\downarrow}D_{2\uparrow}D_{2\downarrow}}}
\Bigl\{
q^4(e^{-i\theta_1}-a_1^2\sqrt{R_{1\uparrow}R_{1\downarrow}})
(e^{-i\theta_2}-a_2^2\sqrt{R_{2\uparrow}R_{2\downarrow}}) +\\+
q^{-4}(\sqrt{R_{1\uparrow}R_{1\downarrow}} - a_1^2e^{i\theta_1})
(\sqrt{R_{2\uparrow}R_{2\downarrow}}- a_2^2e^{i\theta_2}) -
e^{-i\varphi} (\sqrt{R_{1\uparrow}}e^{-i\theta_1 / 2}
-a_1^2\sqrt{R_{1\downarrow}}e^{i\theta_1/2})
(\sqrt{R_{2\uparrow}}e^{-i\theta_2 / 2}
-a_2^2\sqrt{R_{2\downarrow}}e^{i\theta_2/2}) -\\- e^{i\varphi}
(\sqrt{R_{1\downarrow}}e^{-i\theta_1 / 2}
-a_1^2\sqrt{R_{1\uparrow}}e^{i\theta_1/2})
(\sqrt{R_{2\downarrow}}e^{-i\theta_2 / 2}
-a_2^2\sqrt{R_{2\uparrow}}e^{i\theta_2/2}) \Bigr\}. \label{WWW}
\end{multline}
\end{widetext}
Eqs. (\ref{Ieq}), (\ref{PPP}) and (\ref{WWW}) fully determine the
Josephson current in ballistic SNS junctions with spin-active
interfaces and represent the central result of this paper. In the
spin-isotropic limit this result reduces to that obtained in Ref.
\onlinecite{GZ02}.

\section{Josephson current: Specific limits}
The above general expression for the Josephson current contains a
lot of information which can be conveniently
illustrated by considering specific limiting cases.

\subsection{Small spin-mixing angles}
Lets us first analyze the limit of small spin-mixing angles
$\theta_1, \theta_2 \ll 1$. In the tunneling limit
($D_{1\uparrow},D_{1\downarrow},D_{2\uparrow},D_{2\downarrow} \ll
1$) the expression  for the Josephson current defined in Eqs.
(\ref{Ieq}), (\ref{PPP}) and (\ref{WWW}) reduces to a much simpler
result
\begin{multline}
I(\chi)= -2 e \mathcal{A} T \sin\chi \sum_{\omega_n>0}
\int\limits_{|p_\parallel|<p_F} \frac{d^2 p_\parallel}{(2\pi)^2}
\\\times
\Rea \dfrac{4 a_1 a_2
\sqrt{D_{1\uparrow}D_{1\downarrow}D_{2\uparrow}D_{2\downarrow}}}{(1-a_1^2)(1-a_2^2)
(q^4 + q^{-4}-2\cos\varphi)}.
\end{multline}
This expression coincides with the result obtained in
Ref. \onlinecite{GZ02} for SINIS structures with effective
spin-isotropic interface transmissions
$D_1=\sqrt{D_{1\uparrow}D_{1\downarrow}}$ and
$D_2=\sqrt{D_{2\uparrow}D_{2\downarrow}}$.

\begin{figure}
\centerline{\includegraphics[width=80mm]{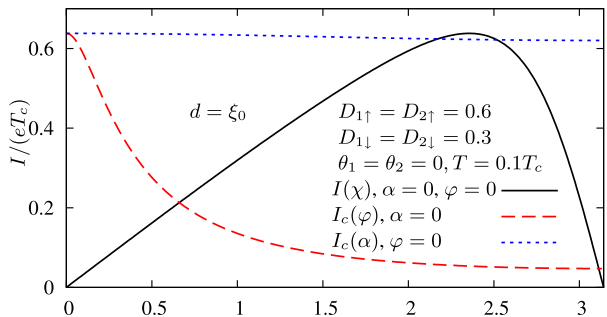}}
\caption{Josephson current across single channel SNS junction with spin-active
interfaces for $d=\xi_0= v_F/(2\pi T_c)$ and zero spin-mixing
angles. The solid curve illustrates the current-phase relation,
while dashed and dotted lines show the dependencies of the
critical current respectively on the kinematic phase $\varphi$ and
on the misorientation angle $\alpha$. } \label{jphi1}
\end{figure}

\begin{figure}
\centerline{\includegraphics[width=80mm]{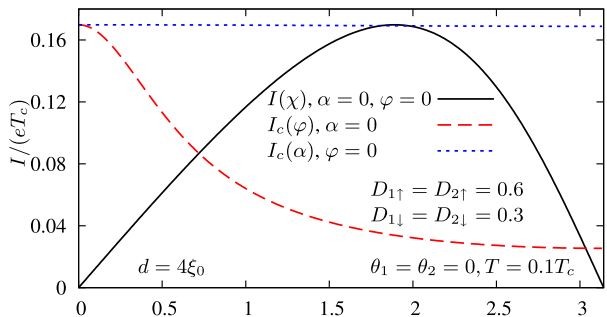}}
\caption{The same as in Fig. 2 for $d=4\xi_0$.} \label{jphi2}
\end{figure}

In the opposite limit of the highly transparent interfaces
$R_{1\uparrow},R_{1\downarrow},R_{2\uparrow},R_{2\downarrow} \ll
1$ and for small spin-mixing angles Eq. \eqref{Ieq} reduces to the
standard expression for the Josephson current in SNS structures
with fully transparent interfaces \cite{IK}. Interestingly, the
Josephson current turns out not to depend on the misorientation
angle $\alpha$ in both limits of weakly and highly transparent
barriers at NS interfaces. At intermediate barrier transmissions
the dependence of the supercurrent on the angle $\alpha$ formally
exists but remains very weak (see Figs. \ref{jphi1} and
\ref{jphi2}). It is also important to point out that no transition
to the $\pi$-junction regime occurs in the limit of the zero
spin-mixing angles.

\subsection{Arbitrary spin-mixing angles}

\begin{figure}
\centerline{\includegraphics[width=80mm]{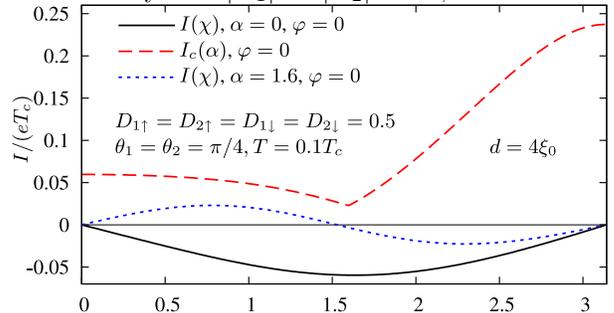}}
\caption{The current-phase relation in single channel SNS junctions with
spin-active interfaces for $d=4\xi_0$ and non-zero spin-mixing
angles. Also the dependence of the critical current$I_c$ on the
misorientation angle $\alpha$ is shown. The transition from $\pi$-
to $0$-junction regime occurs with increasing $\alpha$.}
\label{jphi3}
\end{figure}

In a general case of arbitrary spin-mixing angles the junction
behavior becomes much richer. In particular, the system may now
undergo transitions between $0$- and $\pi$-states driven by
varying kinematic phase $\varphi$, misorientation angle $\alpha$
and temperature. In Fig. \ref{jphi3} we present typical
current-phase relations as well as the critical current dependence
on the misorientation angle $\alpha$. We observe the
$\pi$-to-$0$-junction transition occurs approximately at
$\alpha\approx1.6$ for the chosen values of the parameters.

\subsection{Short junction limit}
An important limiting case is realized provided the thickness of
the normal layer tends to zero $d\to 0$. In practice, this
condition implies $d \ll \xi_0$ off resonance, while at
resonance this inequality turns out to be more stringent
in the limit of small barrier transmissions \cite{GZ02}. In this
short junction limit the general expression for the Josephson
current becomes much simpler. For example, if the scattering matrix at the right interface is proportional to
unity and $|\Delta_1|=|\Delta_2|=\Delta$, we recover the result of Barash and
Bobkova\cite{BB}
\begin{multline}
I=\sum_k \frac{e {\cal D} \Delta\sin \chi}{2\sqrt{1-({\cal R}+{\cal D} \cos\chi)^2}}\left[
\sin\Phi_+\tanh\left(\frac{\Delta\cos\Phi_+}{2T}\right)\right.\\ \left.-\sin\Phi_-
\tanh\left(\frac{\Delta\cos\Phi_-}{2T}\right)\right], \label{bb}
\end{multline}
where ${\cal D}=\sqrt{D_{1\uparrow} D_{1\downarrow}}, {\cal R}=\sqrt{R_{1\uparrow} R_{1\downarrow}}$ and
\begin{equation}
\Phi_\pm=\frac{\theta_1}{2}\pm \frac{1}{2}\arccos\left[ {\cal R}+ {\cal D}\cos\chi \right]. \label{phid}
\end{equation}
Note, that in a general case one has ${\cal R}+{\cal D}\ne 1$. In the case of
non-trivial scattering at the second collinear interface we again arrive at
the Josephson current in the form (\ref{bb}), but obtain more complicated expressions for ${\cal R},{\cal D}, \Phi_\pm$
\begin{widetext}
\begin{multline}
{\cal R}=\frac{\sqrt{R_{1\uparrow }R_{1\downarrow }} \cos \theta_2  + \sqrt{R_{2\uparrow
}R_{2\downarrow }} \cos \theta_ 1 -\sqrt{R_{1\uparrow }R_{2\downarrow
}}\cos((\theta_2-\theta_1)/2-\varphi) -\sqrt{R_{2\uparrow }R_{1\downarrow
}}\cos((\theta_1-\theta_2)/2-\varphi) }{\sqrt{1+R_{1\uparrow } R_{2\uparrow }
-2 \sqrt{R_{1\uparrow }R_{2\uparrow }}\cos((\theta_1+\theta_2)/2-\varphi)}\sqrt{1+R_{1\downarrow
} R_{2\downarrow } -2 \sqrt{R_{1\downarrow }R_{2\downarrow
}}\cos((\theta_1+\theta_2)/2+\varphi)} }
\\ {\cal D}=\frac{\sqrt{D_{1\uparrow }D_{1\downarrow }D_{2\uparrow }D_{2\downarrow }}}{\sqrt{1+R_{1\uparrow } R_{2\uparrow }
-2 \sqrt{R_{1\uparrow }R_{2\uparrow }}\cos((\theta_1+\theta_2)/2-\varphi)}\sqrt{1+R_{1\downarrow
} R_{2\downarrow } -2 \sqrt{R_{1\downarrow }R_{2\downarrow
}}\cos((\theta_1+\theta_2)/2+\varphi)} }.
\end{multline}
As for $\Phi_\pm$, we should also substitute   $\theta_1$ in Eq.(\ref{phid}) by $\theta$ given by
\begin{equation}
\theta=\arctan \frac{\sin (\theta_1+\theta_2)-\sin((\theta_1+\theta_2)/2-\varphi) \sqrt{R_{1\downarrow }
R_{2\downarrow }}-\sin((\theta_1+\theta_2)/2+\varphi) \sqrt{R_{1\uparrow } R_{2\uparrow }}}{\cos(
\theta_1+\theta_2) + \sqrt{R_{1\uparrow }R_{2\uparrow } R_{1\downarrow } R_{2\downarrow
}}-\cos((\theta_1+\theta_2)/2-\varphi)\sqrt{R_{1\downarrow } R_{2\downarrow
}}-\cos((\theta_1+\theta_2)/2+\varphi)\sqrt{R_{1\uparrow }R_{2\uparrow }}}.
\label{tet}
\end{equation}
\end{widetext}

\begin{figure}
\includegraphics[width=80mm]{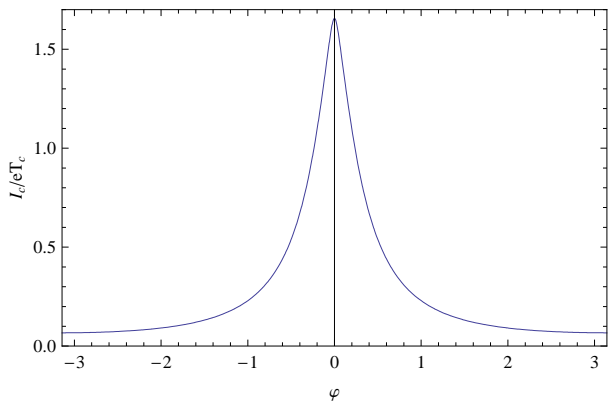}
\caption{The critical Josephson current for a short collinear
($\alpha=0$) single-channel SNS junction with $D_{1\uparrow}=D_{2\uparrow}=0.6,
D_{1\downarrow}=D_{2\downarrow}=0.3$ and $\theta_{1,2}=0$. Here we
have set $T=0.1 T_c$.} \label{shortj1}
\end{figure}
The last expression applies provided its denominator takes positive values,
otherwise $\pi$ should be added to to the right-hand side of Eq. (\ref{tet}).

\begin{figure}
\includegraphics[width=80mm]{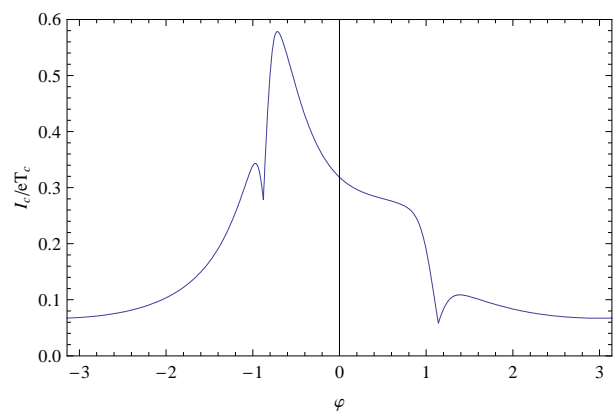}
\caption{The same as in Fig. \ref{shortj1} for
$\theta_1=\theta_2=\pi/4$.} \label{shortj2}
\end{figure}

\begin{figure}
\includegraphics[width=80mm]{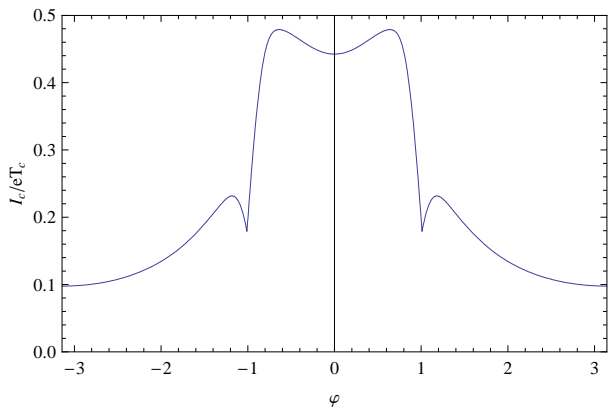}
\caption{The same as in Figs. \ref{shortj1} and \ref{shortj2} for
$D_{1\uparrow}=D_{2\uparrow}=D_{1\downarrow}=D_{2\downarrow}=0.5$
and $\theta_{1,2}=\pi/4$.} \label{shortj3}
\end{figure}

Current experimental techniques permit measuring Josephson current
through junctions with few conducting channels formed, e.g. by
single wall carbon nanotubes \cite{Nano,Lindelof,Basel} or
metallo-fullerenes \cite{Bouchiat}.  In this case one can
effectively tune the Josephson current by applying the gate
voltage to the junction. This kind of measurements is essentially
equivalent to detecting the $\varphi$-dependence of the
supercurrent. Changing the gate voltage (or the phase $\varphi$)
one can drive the system towards resonance, thereby reaching
substantial enhancement of the critical current. This resonance
effect was experimentally demonstrated, e.g., in Ref.
\onlinecite{Nano}. In the case of SNS junctions with spin-active
interfaces considered here the form of the resonant current peak
becomes much more complicated. The corresponding examples are
presented in Figs. \ref{shortj1}, \ref{shortj2} and \ref{shortj3}.

We observe that at zero spin-mixing angles $\theta_{1,2}$ the form
of the resonance peak is essentially identical to that for
non-magnetic SNS junctions \cite{GZ02}, as demonstrated in Fig.
\ref{shortj1}. The situation changes dramatically for non-zero
values of the spin-mixing angles. In this case the resonant
current peak in general becomes asymmetric and, on top of that, an
additional dip structure emerges, see Fig. \ref{shortj2}. It is
important to emphasize that the current peak asymmetry is due to
different spin-up and spin-down transmission values $D_\uparrow$
and $D_\downarrow$. For equal values $D_\uparrow$ and
$D_\downarrow$ the symmetry of the current peak is restored,
however the peak now becomes split into two and also an additional
dip structure remains as long as spin-mixing angles differ from
zero, see Fig. \ref{shortj3}. We also note that in the case of small
barrier transmissions the current resonances are reached at the points
$\cos[(\theta_1+\theta_2)/2\pm\varphi]=1$.

\begin{figure}
\includegraphics[width=80mm]{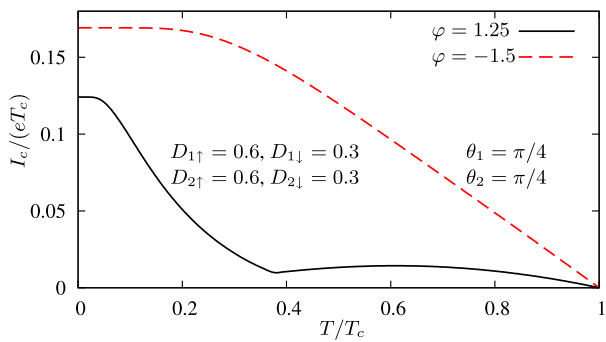}
\caption{The critical Josephson current for short collinear
($\alpha=0$) single channel SNS junctions as a function of temperature.}
\label{tdep}
\end{figure}

The temperature dependence of the Josephson critical current for
short SNS junctions is shown in the Fig. \ref{tdep}. It turns out
that the temperature dependence can be non-monotonous even in the
limit $d\ll \xi_0$. Note, however, that this non-monotonous behavior
may occur only in the narrow range of $\varphi$ in the vicinity of
the local minimum of the $I_c(\varphi)$-dependence, cf. Figs.
\ref{shortj2} and \ref{shortj3}.

\subsection{Long junctions at non-zero temperatures}

\begin{figure}
\includegraphics[width=8cm]{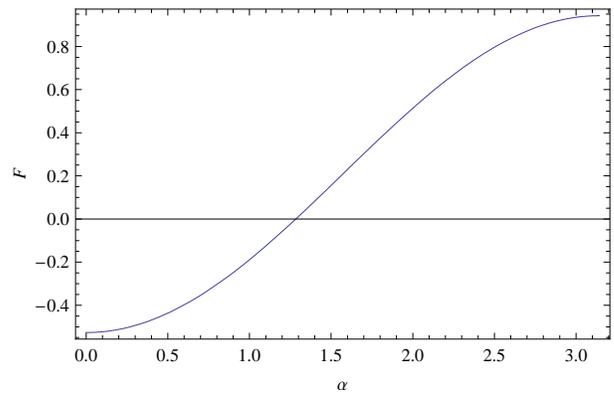}
\caption{The function $F$ from Eq. (\ref{ff}) for identical
interfaces with $\sqrt{R_{\uparrow}R_{\downarrow}}=0.9$ and
$\theta_{1,2}=2.0$} \label{ljfig}
\end{figure}

In the limit of long junctions with $d\gg v_F/T$ the result for
the Josephson current simplifies dramatically and reads
\begin{multline}
I=8eT\sin\chi \mathcal{A}\int\limits_{|p_\parallel|<p_F} \frac{d^2
p_\parallel}{(2\pi)^2} \sqrt{D_{1\uparrow} D_{1\downarrow}
D_{2\uparrow} D_{2\downarrow} } \times\\\times \exp(-2\pi T d/v_{F
x})  F, \label{ff}
\end{multline}
where
\begin{multline}
F= \cos^2\frac{\alpha}{2}\,\mathrm{Re}
\left[\frac{1}{\left(e^{i\theta_1}+\sqrt{R_{1\uparrow}R_{1\downarrow}}\right)
\left(e^{i\theta_2}+\sqrt{R_{2\uparrow}R_{2\downarrow}}\right)}\right]
+\\+  \sin^2\frac{\alpha}{2}\,\mathrm{Re}
\left[\frac{1}{\left(e^{-i\theta_1}+\sqrt{R_{1\uparrow}R_{1\downarrow}}\right)
\left(e^{i\theta_2}+\sqrt{R_{2\uparrow}R_{2\downarrow}}\right)}\right].\nonumber
\end{multline}
Note that even in this limit the above expression retains
non-trivial features related to the presence of spin-active
interfaces. In particular, for large enough values of the
$\theta$-angles the $\pi$-junction behavior can be realized. Also,
the transition between $0$- and $\pi$-junction regimes can occur
depending on the value $\alpha$. For example, in the case of
identical weakly transmitting interfaces the $\pi$-junction regime
is realized in the range $\pi/2<\theta_{1,2}<3\pi/2 $ . This
behavior is demonstrated in Fig. \ref{ljfig}. We observe that for
smaller values of $\alpha$ the function $F(\alpha )$ takes
negative values which corresponds to the $\pi$-junction regime. In
contrast, for larger $\alpha$ the function $F(\alpha )$ becomes
positive and the standard 0-junction regime is realized.

\section{Concluding remarks}

In this paper we developed a general microscopic theory of dc Josephson effect
in hybrid SNS structures with ballistic electrodes and spin-active NS
interfaces. We evaluated  the spectrum of Andreev levels (\ref{peq})
and derived a direct and simple relation between this spectrum and
the Josephson current in such systems. The latter current can be expressed in
a very compact general form (\ref{Ieq}) which contains complete information
about non-trivial interplay between Andreev reflection and spin-dependent
scattering at NS interfaces. Our analysis revealed a rich structure of
properties realized in various physical limits. These properties
-- along with ``usual'' dependencies of the supercurrent
on the junction size and temperature -- also crucially depend on
spin-dependent barrier transmissions, spin-mixing angles, relative
magnetization orientation of interfaces as well as on the
kinematic phase of scattered electrons.

In particular, we studied the
current-phase relation in our system and demonstrated a broad range of
different regimes including the presence of a $\pi$-junction state which
may only be possible in the case of non-zero spin mixing angles.
We also analyzed the effect of resonant enhancement of the critical current
which mainly occurs in nanojunctions with few conducting channels
driven by an externally applied gate voltage. We found that -- in
contrast to the non-magnetic case -- spin-dependent scattering may dramatically
modify both the magnitude and the shape
of a resonant peak, e.g., making it asymmetric and/or providing an additional
structure of dips, cf. Figs. 5, 6 and 7. We believe that these and other our
predictions can be directly tested in experiments with hybrid
superconducting-normal proximity structures containing, e.g. thin
ferromagnetic layers. We also anticipate that further theoretical activity in
the field might help to discover new exciting properties of the systems under consideration.

\centerline{\bf Acknowledgments}

\vspace{0.5cm}

 This work was supported in part by RFBR grant 06-02-17459. M.S.K. also
acknowledges support from the Dynasty Foundation.

\appendix
\section{Green functions and supercurrent}
\label{green} In order to derive the general expression for the
Josephson current let us evaluate the Green functions $G(x,x')$
for our system. Inside the normal metal and for $d_1<x<x'<d_2$ we
have
\begin{multline}
G(x,x')=
\begin{pmatrix}
e^{i\varepsilon(x-d_1)/v_{Fx}} B_1^-(x') \\
e^{-i\varepsilon(x-d_1)/v_{Fx}} B_2^-(x') \\
\end{pmatrix}
e^{ip_{Fx}(x-d_1)}
+\\+
\begin{pmatrix}
e^{-i\varepsilon(x-d_1)/v_{Fx}} B_3^-(x') \\
e^{i\varepsilon(x-d_1)/v_{Fx}} B_4^-(x') \\
\end{pmatrix}
e^{-ip_{Fx}(x-d_1)}, \label{A1}
\end{multline}
while for $d_1<x'<x<d_2$ we write
\begin{multline}
G(x,x')=
\begin{pmatrix}
e^{i\varepsilon(x-d_1)/v_{Fx}} B_1^+(x') \\
e^{-i\varepsilon(x-d_1)/v_{Fx}} B_2^+(x') \\
\end{pmatrix}
e^{ip_{Fx}(x-d_1)}
+\\+
\begin{pmatrix}
e^{-i\varepsilon(x-d_1)/v_{Fx}} B_3^+(x') \\
e^{i\varepsilon(x-d_1)/v_{Fx}} B_4^+(x') \\
\end{pmatrix} e^{-ip_{Fx}(x-d_1)},
\label{A2}
\end{multline}
where $B^{\pm}_i(x)$ are $2\times4$
matrices. The coefficients $B^{\pm}_i(x)$ are determined from the
relations
\begin{gather}
G(x,x+0)-G(x,x-0)=0,
\label{Geq1}
\\
\left.\dfrac{\partial G(x,x')}{\partial x}\right|_{x'=x-0}-
\left.\dfrac{\partial G(x,x')}{\partial x}\right|_{x'=x+0}=
2 m \tau_3,
\label{Geq2}
\end{gather}
and interface boundary conditions (cf. Eqs. (\ref{k2rel}),
(\ref{k2rel}))
\begin{gather}
\begin{pmatrix}
B_1^-(x) \\
B_2^-(x) \\
\end{pmatrix}
=
K_1
\begin{pmatrix}
B_3^-(x) \\
B_4^-(x)
\end{pmatrix},
\\
\begin{pmatrix}
B_3^+(x) \\
B_4^+(x) \\
\end{pmatrix}
=
e^{-i\varphi}
Q K_2 Q
\begin{pmatrix}
B_1^+(x) \\
B_2^+(x)
\end{pmatrix}.
\end{gather}
Eqs. \eqref{Geq1} and \eqref{Geq2} yield
\begin{multline}
\begin{pmatrix}
B_1^-(x) \\
B_2^-(x)
\end{pmatrix}
=
\begin{pmatrix}
B_1^+(x) \\
B_2^+(x)
\end{pmatrix}
+\\+
\dfrac{i}{v_{Fx}}
\begin{pmatrix}
T_1 e^{-i\varepsilon(x-d_1)/v_{Fx}}\\
- T_2 e^{i\varepsilon(x-d_1)/v_{Fx}}\\
\end{pmatrix}
e^{-i p_{Fx}(x-d_1)},
\end{multline}
\begin{multline}
\begin{pmatrix}
B_3^-(x) \\
B_4^-(x)
\end{pmatrix}
=
\begin{pmatrix}
B_3^+(x) \\
B_4^+(x)
\end{pmatrix}
-\\-
\dfrac{i}{v_{Fx}}
\begin{pmatrix}
T_1 e^{i\varepsilon(x-d_1)/v_{Fx}}\\
-T_2 e^{-i\varepsilon(x-d_1)/v_{Fx}}\\
\end{pmatrix}
e^{i p_{Fx}(x-d_1)},
\end{multline}
where we introduced the matrices
\begin{equation}
T_1=
\begin{pmatrix}
1 & 0 & 0 & 0 \\
0 & 1 & 0 & 0 \\
\end{pmatrix},
\quad
T_2=
\begin{pmatrix}
0 & 0 & 1 & 0 \\
0 & 0 & 0 & 1 \\
\end{pmatrix},
\end{equation}
Finally we obtain
\begin{widetext}
\begin{multline}
\begin{pmatrix}
B_1^+(x) \\
B_2^+(x)
\end{pmatrix}=
-\dfrac{i}{v_{Fx}}
\Biggl\{
\left(1-e^{-i\varphi}K_1 Q K_2 Q \right)^{-1}K_1
\begin{pmatrix}
T_1 e^{i\varepsilon(x-d_1)/v_{Fx}}\\
-T_2 e^{-i\varepsilon(x-d_1)/v_{Fx}}\\
\end{pmatrix}
e^{i p_{Fx}(x-d_1)}
+\\+
\left(1-e^{-i\varphi}K_1 Q K_2 Q\right)^{-1}
\begin{pmatrix}
T_1 e^{-i\varepsilon(x-d_1)/v_{Fx}}\\
- T_2 e^{i\varepsilon(x-d_1)/v_{Fx}}\\
\end{pmatrix}
e^{-i p_{Fx}(x-d_1)} \Biggr\}, \label{A10}
\end{multline}
\begin{multline}
\begin{pmatrix}
B_3^+(x) \\
B_4^+(x)
\end{pmatrix}=
-\dfrac{i}{v_{Fx}}
\Biggl\{
\left(1-e^{-i\varphi}Q K_2 Q K_1 \right)^{-1}
e^{-i\varphi} Q K_2 Q K_1
\begin{pmatrix}
T_1 e^{i\varepsilon(x-d_1)/v_{Fx}}\\
-T_2 e^{-i\varepsilon(x-d_1)/v_{Fx}}\\
\end{pmatrix}
e^{i p_{Fx}(x-d_1)}
+\\+
\left(1-e^{i\varphi}Q K_2 Q K_1\right)^{-1}
e^{-i\varphi} Q K_2 Q
\begin{pmatrix}
T_1 e^{-i\varepsilon(x-d_1)/v_{Fx}}\\
- T_2 e^{i\varepsilon(x-d_1)/v_{Fx}}\\
\end{pmatrix}
e^{-i p_{Fx}(x-d_1)}
\Biggr\}
\label{A11}
\end{multline}
\end{widetext}
Now we are ready to evaluate the Josephson current. Combining Eqs.
(\ref{A1}), (\ref{A2}) and (\ref{A10}), (\ref{A11}) with the
general formula for the current
\begin{multline}
I=\dfrac{ieT\mathcal{A}}{4 m}
\sum_{\omega_n>0}
\int\limits_{|p_\parallel|<p_F}
\frac{d^2 p_\parallel}{(2\pi)^2}
\times\\\times
\Sp\left[ (1+\tau_3)(\nabla_{x'}-\nabla_{x})_{x'\rightarrow x}
G(x,x',i\omega_n)\right]
\end{multline}
we arrive at the following expression
\begin{multline}
I(\chi)=
-e \mathcal{A} T
\sum_{\omega_n>0}
\int\limits_{|p_\parallel|<p_F}
\frac{d^2 p_\parallel}{(2\pi)^2}
\Img
\Sp\biggl[(1+\tau_3)
\times\\
\left\{\left(1-e^{-i\varphi}K_1 Q K_2 Q \right)^{-1}-
\left(1-e^{-i\varphi} Q K_2 Q K_1 \right)^{-1}\right\}
\biggr].
\label{Ieqgreen}
\end{multline}
Making use of the identities
\begin{gather}
\dfrac{\partial K_1}{\partial \chi}=
-\dfrac{i}{4}(\tau_3 K_1 - K_1\tau_3),
\\
\dfrac{\partial K_2}{\partial \chi}=
\dfrac{i}{4}(\tau_3 K_2 - K_2\tau_3),
\\
\dfrac{\partial \det| A(t)|}{\partial t}=
\Sp[A^{-1}(t)A'_t(t)]\det| A(t)|
\end{gather}
we verify the full equivalence of Eqs. \eqref{Ieqgreen} and
\eqref{Ieq}.

\section{Channel-mixing scattering}
Within our analysis we assumed that the electron momentum parallel
to spin-active NS interfaces is conserved during scattering at
such interfaces. This condition is equivalent to the absence of
mixing between different transmission channels. The purpose of
this appendix is to demonstrate that our approach can be
straightforwardly generalized to the case of channel-mixing
scattering.

Assume that there exist ${\cal N}_N$ transmission channels in a
normal metal layer, that can be scattered into ${\cal N}_{S1}$ and
$ {\cal N}_{S2}$ channels respectively in the left and right
superconductors. The current amplitudes related via the scattering
matrix will be numbered in the following order: spin-up current
amplitudes at the left-hand side, spin-down current amplitudes at
the left-hand side (ordered in the same way as spin-up amplitudes,
i.e. for the same order of scattering channels), spin-up current
amplitudes at the right-hand side and spin-down current amplitudes
at the right-hand side. Thus, the block $S_{11}$ describing
electron reflection back into the left superconductor is given by
the $2{\cal N}_{S1}\times 2{\cal N}_{S1}$ matrix and $S_{1'1'}$
block which accounts for electron reflection into the normal layer
is presented by the $2{\cal N}_N\times 2{\cal N}_N$ matrix. The
blocks $S_{11'}, S_{1'1}$ are rectangular with dimensions $2 {\cal
N}_{S1}\times 2 {\cal N}_{N}$ and $2 {\cal N}_{N}\times 2 {\cal
N}_{S1}$ correspondingly. Hence, the total dimensions of the unitary
matrix ${\cal S}_1$ are $2({\cal N}_{S1}+{\cal N}_N)\times 2({\cal
N}_{S1}+{\cal N}_N)$.

The dimensions of blocks $S_{22}$ , $S_{2'2'}$, $S_{22'}$, and
$S_{2'2}$ are $2{\cal N}_N\times 2{\cal N}_N$,  $2{\cal
N}_{S2}\times 2{\cal N}_{S2}$ , $2 {\cal N}_{N}\times 2 {\cal
N}_{S2}$, and $2 {\cal N}_{S2}\times 2 {\cal N}_{N}$. The blocks
corresponding to the hole scattering matrices
$\underline{S}_{1,2}$ are characterized by the same dimensions.
From the unitary scattering matrices $S_1$ and $\underline{S}_1$ we
construct the $4{\cal N}_N\times 4{\cal N}_N$ matrix
\begin{equation}
\check M_1=\left( \begin{array}{cc} \hat e_1& \hat f_1 \\ \hat g_1
& \hat h_1\end{array}\right).
\end{equation}
The matrices $\hat e_1, \hat f_1, \hat g_1, \hat h_1$ have
dimensions $2{\cal N}_N\times 2{\cal N}_N$ and are defined by
\begin{widetext}
\begin{eqnarray}
&& \hat e_1
=\left(1-\underline{S}_{1'1'}\underline{S}^+_{1'1'}\right)^{-1}
\underline{S}_{1'1}\left( -i a_1\underline{S}_{11}^+\hat
\sigma_{y1} S_{11}+ia_1^{-1}\hat \sigma_{y1}\right)\left(
1-S_{11}^+S_{11}\right)^{-1}  S^+_{1'1},
\\ && \hat f_1=\left(1-\underline{S}_{1'1'}\underline{S}^+_{1'1'}\right)^{-1}
\underline{S}_{1'1}\left(-i a_1\underline{S}_{11}^+\hat\sigma_{y1}+
i a_1^{-1}\hat\sigma_{y1} S_{11}^+\right) \left( 1-S_{11}S_{11}^+\right)^{-1} S_{11'},\nonumber
\\ &&  \hat g_1=-\left(1-\underline{S}_{1'1'}^+\underline{S}_{1'1'}\right)^{-1}\underline{S}_{11'}^+
\left( -i a_1\hat\sigma_{y1} S_{11}+i
a_1^{-1}\underline{S}_{11}\hat\sigma_{y1}\right) \left(
1-S_{11}^+S_{11}\right)^{-1}  S^+_{1'1},\nonumber
\\ && \hat h_1=- \left(1-\underline{S}_{1'1'}^+\underline{S}_{1'1'}\right)^{-1}
\underline{S}_{11'}^+ \left( -i a_1\hat\sigma_{y1}+ia_1^{-1}\underline{S}_{11}\hat\sigma_{y1}
S_{11}^+\right)\left( 1-S_{11}S_{11}^+\right)^{-1} S_{11'}.\nonumber
\end{eqnarray}

Similarly, the unitary matrices $S_{2},\underline{S}_2$ describing
electron scattering at the right interface are defined by
\begin{eqnarray}
&& \hat e_2
=\left(1-\underline{S}_{22}\underline{S}^+_{22}\right)^{-1}
\underline{S}_{22'}\left(  -i a_2\underline{S}_{2'2'}^+\hat
\sigma_{y2} S_{2'2'}+ia_2^{-1}\hat \sigma_{y2}\right)\left(
1-S_{2'2'}^+S_{2'2'}\right)^{-1}  S^+_{22'},
\\ && \hat f_2=\left(1-\underline{S}_{22}\underline{S}^+_{22}\right)^{-1}
\underline{S}_{22'}\left(-i a_2\underline{S}_{2'2'}^+\hat\sigma_{y2}+
i a_2^{-1}\hat\sigma_{y2} S_{2'2'}^+\right) \left( 1-S_{2'2'}S_{2'2'}^+\right)^{-1} S_{2'2},\nonumber
\\ &&  \hat g_2=-\left(1-\underline{S}_{22}^+\underline{S}_{22}\right)^{-1}\underline{S}_{2'2}^+
\left(-i a_2\hat\sigma_{y2} S_{2'2'}+i
a_2^{-1}\underline{S}_{2'2'}\hat\sigma_{y2}\right) \left(
1-S_{2'2'}^+S_{2'2'}\right)^{-1}  S^+_{22'},\nonumber
\\ && \hat h_2=- \left(1-\underline{S}_{22}^+\underline{S}_{22}\right)^{-1}
\underline{S}_{2'2}^+ \left( -i a_2\hat\sigma_{y2}+i a_2^{-1}\underline{S}_{2'2'}\hat\sigma_{y2}
S_{2'2'}^+\right)\left( 1-S_{2'2'}S_{2'2'}^+\right)^{-1} S_{2'2}.\nonumber
\end{eqnarray}
\end{widetext}
Comparing these expressions we note that the matrices which account for the
second interface are obtained from ones for the first interface by
substituting the indices $1\to 2$ and, in addition, by
interchanging the indices with and without the primes. The
matrices $\hat\sigma_{y1,2}$ have the dimensions $2{\cal N}_{
S1,2}\times 2{\cal N}_{ S1,2}$ and the following structure
\begin{equation}
\hat\sigma_y=i\left(\begin{array}{cc} 0& -\hat 1 \\ \hat 1 & 0
\end{array} \right),
\end{equation}
where $\hat 1$ is the unity matrix. The dimensions of this matrix
are ${\cal N}_{ S1}\times {\cal N}_{ S1}$ in the expression for
$\hat\sigma_{y1}$ and ${\cal N}_{ S2}\times {\cal N}_{ S2}$ in the
expression for $\hat \sigma_{y2}$. As before, we define
$a_{1,2}=i(\sqrt{\omega_n^2+|\Delta_{1,2}|^2}-\omega_n)/|\Delta_{1,2}|$.

In order to establish the general expression for the current in
our system it is necessary to generalize the parameter $q$
(defined in Eq. (\ref{qdef})) to the current matrix form as well
as to incorporate kinematic phases for different transmissions
channels into the matrix structure considered here. This task is
accomplished by defining the following $4{\cal N}_N\times 4{\cal
N}_N$ matrices
\begin{equation}
\check Q=\left( \begin{array}{cc} \hat Q & 0 \\ 0 & \hat
Q^{-1}\end{array}\right),\quad \check Z=\left( \begin{array}{cc} 0
& \hat Z^{-1}\\ \hat Z & 0\end{array}\right).
\end{equation}
The diagonal  $2{\cal N}_N\times 2{\cal N}_N$ matrix $\hat Q$
reads
\begin{equation}
\hat Q=\left(\begin{array}{cccccc} q_1& & & & &\\  & \ddots& & &
&\\ & &q_{ {\cal N}_N} & & &\\  & & &q_1 & &\\ & & & &\ddots &\\ &
& & & &q_{ {\cal N}_N} \end{array}\right).
\end{equation}
Here the indices label the conducting channels and
we defined $q_k=\exp(\omega_n d/2v_{Fk})$ with $v_{Fk}$ standing for the
$x$-component of the Fermi velocity for a given conductance
channel. This structure of the matrix $\hat Q$ stems from the
channel ordering convention adopted here. We also point out that
the Fermi velocities for spin-up and spin-down electrons coincide
for each transmission channel since both normal metal and
superconducting electrodes are assumed to be non-magnetic.

Similarly, the matrix $\hat Z$ has the form
\begin{equation}
    \hat Z=\left(\begin{array}{cccccc} z_1& & & & &\\  & \ddots& & & &\\ & &z_{ {\cal N}_N}
& & &\\  & & &z_1 & &\\ & & & &\ddots &\\ & & & & &z_{ {\cal N}_N}
\end{array}\right),
\end{equation}
where $z_k=\exp(ip_{Fk}d)$.

With the aid of these matrices we derive the contribution to the
Josephson current $I_+$ defined by positive Matsubara frequencies
as
\begin{equation}
I_+=-eT\sum_{\omega_n>0}\frac{d}{d\chi} \ln P(\chi),
\end{equation}
where
\begin{equation}
P(\chi)=\det\left| e^{i\chi/2}\check Z\check
M_{Q1}-e^{-i\chi/2}\check M_{Q2}\check Z\right|.
\end{equation}
The matrices $\check M_{Q1,2}$ are defined as
\begin{equation}
\check M_{Q1,2}=\check Q \check M_{1,2}\check Q.
\end{equation}
The negative Matsubara frequencies contribution $I_-$ is obtained
from $I_+$ by the substitution $q_k\rightarrow q_k^{-1}$ and
$a_{1,2}\rightarrow -a_{1,2}^{-1}$. Provided the symmetry
defined in Eq. (\ref{srel}) holds, the relation $I_-=I_+$ follows immediately.

In the absence of channel mixing the function $P(\chi)$ is
factorized into the product of contributions from independent
channels. Accordingly, the supercurrent is expressed as a sum over
these channels. In this case the function $P(\chi)$ coincides with that derived
above in this paper and the corresponding expression for the
current reduces to that defined in Eqs. (\ref{Ieq}), (\ref{PPP})
and (\ref{WWW}).

\end{document}